\documentclass[aps,prl,showpacs,twocolumn,superscriptaddress,letter]{revtex4}

\usepackage{epsfig}
\usepackage{amsmath,amssymb,wasysym,epsfig,capt-of,ifthen,calc}
\usepackage{bbm}
\usepackage{latexsym} 
\usepackage{setspace} 
\usepackage{array} 

\usepackage{delarray} 
\usepackage{afterpage} \usepackage{graphicx}
\usepackage{dcolumn}
\usepackage{bm}
\usepackage{float} \usepackage{supertabular} \usepackage{longtable}
\newcommand{\be}{\begin{equation}} \newcommand{\ee}{\end{equation}}
\newcommand{\bea}{\begin{eqnarray}} \newcommand{\eea}{\end{eqnarray}}

\usepackage{amsmath}\bibstyle{apsrev}

\begin{document}
\title{Irreversible Aggregation  and Network Renormalization}

\author{Seung-Woo Son} \affiliation{Complexity Science Group, University of Calgary, Calgary T2N 1N4, Canada}

\author{Golnoosh Bizhani} \affiliation{Complexity Science Group, University of Calgary, Calgary T2N 1N4, Canada}

\author{Claire Christensen} \affiliation{Complexity Science Group, University of Calgary, Calgary T2N 1N4, Canada}

\author{Peter Grassberger} \affiliation{Complexity Science Group, University of Calgary, Calgary T2N 1N4, Canada} \affiliation{FZ J\"ulich, D-52425 J\"ulich, Germany}

\author{Maya Paczuski} \affiliation{Complexity Science Group, University of Calgary, Calgary T2N 1N4, Canada}

\date{\today}

\begin{abstract}
Irreversible aggregation is revisited in view of
recent work on renormalization of complex networks. Its scaling laws
and phase transitions are related to percolation transitions seen
in the latter. We illustrate our points by giving the complete solution for the
probability to find any given state in an aggregation process
$(k+1)X\to X$, given a fixed number of unit mass particles in the
initial state. Exactly the same probability distributions and scaling are found in one
dimensional systems (a trivial network) and well-mixed solutions.  This reveals that scaling laws found in renormalization of complex networks do not prove that they are self-similar.
\end{abstract}
\maketitle

Droplets beget rain,  goblets coagulate to make butter or cream,
and dust particles stick together to form aggregates  that can
eventually coalesce into planets. At the microscopic level,
irreversible aggregation of atoms and molecules creates many
familiar forms of matter such as aerosols, colloids, gels,
suspensions, clusters and solids~\cite{zangwill2001statistical}.
Almost a century ago, Smoluchowski proposed a theory based on rate
equations to describe processes governed by
diffusion, collision and irreversible merging of
aggregates~\cite{rate_equation}. The theory predicts how many
small and large clusters exist at any given time and yields a mass
distribution that depends on certain details such as the initial
conditions, reactions present, relative rates, the presence or absence of spatial structure, etc. A key
interest to physicists  has been to
derive scaling laws that characterize different
universality classes~\cite[and references therein]{leyvraz}.

By contrast, wide interest in complex networks~\cite{Strogatz2001,Albert2002,Dorogovtsev2002,Newman2003}   has emerged recently.  Vast applications to
physics, computer science, biology, and sociology~\cite[and
references therein]{Dorogovtsev2008,Barabasi2009,Barabasi2011}
continue to be vigorously investigated. An important question is
whether or not
complex networks exhibit self-similarity at different length
scales and if they can be grouped into universality classes on that basis. Renormalization schemes for  networks were
proposed~\cite{CSong2005,KGoh2006,JKim2007,Rozenfeld2010} to
address this question.   Scaling of the mass or degree distribution of the renormalized nodes  was used to argue that many complex networks are self-similar. The semi-sequential renormalization group
(RG) flow underlying the box covering
of~\cite{CSong2005,KGoh2006,JKim2007,Rozenfeld2010} was studied
carefully in~\cite{Radicchi2008,Radicchi2009}, where it was found
that scaling laws may be related to an ``RG fixed point'' which
was observed for a wide variety of networks.  A convenient, fully sequential scheme called
random sequential renormalization (RSR) was introduced~\cite{bizhani2010}.   At each RSR step, one
node is selected at random, and all nodes within a fixed distance $\ell$
of it are replaced by a single super-node.

We point out  a simple mapping  between RSR and irreversible
aggregation on any graph. Hence any conclusion drawn for one
process holds also for the other.  Indeed, a local coarse-graining
step to produce a new super-node represents one aggregation event,
where a `molecule' aggregates with all its  neighbors within
distance $\ell$ to produce a new cluster.  Exact analysis  in one
dimension reveals that even this trivial network exhibits scaling
laws for the cluster mass distribution under RSR -- with exponents
that depend on $\ell$. Consequently, and somewhat
counter-intuitively, self-similarity observed in RSR and similar
network renormalization schemes cannot be used to prove that
complex networks are themselves self-similar.  Instead scaling
laws arise due to a percolation transition in irreversible
aggregation.

The correspondence between aggregation and renormalization is
relevant for any model with stochastic coarse-graining of a
network.  For instance, the theory of space and time
``Graphity"~\cite{Konopka2006,Hamma2010}, based on loop quantum
gravity,  involves a stochastic coarsening similar (albeit more
structured) to RSR.  Hence the critical point of aggregation may
also be relevant  in that and related cases. The breakdown of
conventional universality, where critical exponents depend on the
microscopic scale of  coarse-graining, $\ell$,  seems to present a
dilemma for theories based  on stochastic coarse-graining of a
network to arrive at e.g. a universal large scale theory of
gravity.

\begin{figure}[]
\includegraphics[width=0.62\columnwidth]{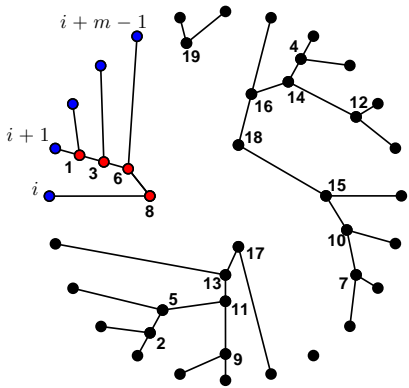}
\caption{(Color online) Illustration of  aggregation on a ring
with $k=1$, $N_0=24$, and $N=5$. The tree in color corresponds to
a cluster of mass $m=5$. It has five leaves (blue) and four
internal nodes (red). Its leaves start at site $i$ and end at site
$i+m-1$. The numbers beside internal nodes correspond to the time
when coalescence occurs.}        \label{fig0}
\vskip -.2in

\end{figure}


In order to demonstrate these points, here we consider
irreversible aggregation $(k+1)X \to X$, where a randomly picked
cluster coalesces with $k$ neighbors. For even $k=2\ell$ this
corresponds precisely to RSR on a 1-d chain with coarsening range
$\ell$. The mass of the newly formed cluster is the sum of the
$(k+1)$ masses.  We assume that the `target'  cluster is picked
with uniform probability from all clusters. Other choices will be
discussed in~\cite{SWSon2011}.

Let us start with the model defined on a ring, i.e., with periodic
boundary conditions.  Initially, $N_0$ sites labelled by $i\in
[1,...N_0]$ are each occupied by a particle of mass $m=1$. Time
can be either discrete or continuous, but we demand that two
events never happen simultaneously.  Hence events, ranked by
increasing time,  are denoted by positive integer values $t$.  For
each event, particles coagulate to form clusters of mass $m>1$.
More precisely, an event consists of picking a random cluster with
uniform probability and joining it with  $k$ clusters to its
immediate right, using periodic boundary conditions.  For $k$
even, the same results are found if we aggregate clusters
symmetrically. After $t$ events, $N_t = N_0-kt$ clusters exist.
Our main result is the probability to find any sequence of
adjacent cluster masses $p^{N_0}_{N_t}(m_1,m_2\dots m_{N_t})$ --
where a cluster of mass $m_1$ is followed by a cluster of mass
$m_2$, etc., moving clockwise (see Fig.~\ref{fig0}). we start with
the single cluster mass probability.

Cluster masses are restricted to  $
    m \equiv 1$ (mod  $k$).
Defining $m-1 =ks$,  the integer $s$ is the number of events
needed to make  the cluster of mass $m$.  As depicted in
Fig.~\ref{fig0}, we can represent any realization of the process
by a forest of $N_t$ rooted trees with $N_0$ leaves and $t$
internal nodes.  Each tree $\alpha$ has $s_{\alpha}$ internal
nodes, with $\sum_{\alpha} s_{\alpha} = t$.  We simplify the
notation by $N$ for $N_t$.

Let $\pi_N^{N_0}(m)$  denote, for fixed $k$ (the dependence on $k$
is not written explicitly in the following), the probability that
a cluster of mass $m$ has its left-most member at site $i\in
[1,N_0]$ after $t$ events. The probability that any of the $N$
clusters picked at random has mass $m$ is then \be
   p_N^{N_0}(m) = {N_0\over N} \; \pi_N^{N_0}(m) \quad ,        \label{pi}
\ee because there are $N_0$ choices for  $i$ and the chance to
pick that particular cluster, given that it exists, is $1/N$.
Since events occur completely at random, each {\it history} occurs
with equal probability.  The term `history' refers to a fixed
forest, which includes a fixed temporal order of events. Thus
$\pi_N^{N_0}(m)$ is equal to the number of histories leading to a
final configuration with a cluster of mass $m$ starting at
position $i$, divided by all possible histories leading to $N$
clusters. The latter is equal to \be
   n_{\rm hist,tot} = N_0 \times (N_0-k) \times \ldots (N+k),
\ee where each of the $t$ factors equals the number of choices for
the next event. Using  Pochhammer $k-$symbols  or, equivalently,
generalized rising factorials~\cite{normand, ddiaz,pitman,
kingman1982genealogy}, this can be written as $n_{\rm hist,tot} =
\mathbf{(}N+k\mathbf{)}_{t,k}$. Similarly, the number of histories
leading to a cluster of size $m$ starting at a fixed position $i$
is \be
   n_{\rm hist,cluster} = (m-k)(m-2k)\times \ldots 1 = \mathbf{(}1\mathbf{)}_{s,k}   \label{hist_clust}
\ee
and the number of histories for the remaining $N-1$ clusters is
\bea
   n_{\rm hist,rest} &=& (N_0-m-k)(N_0-m-2k)\times \ldots (N-1) \nonumber \\
                     &=& \mathbf{(}N-1\mathbf{)}_{t-s,k}\quad .                         \label{hist_rest}
\eea
So far  we have not included the number of choices associated with  different
time orderings for the $s$ events  in the cluster and  $(t-s)$ events in the rest of
the forest.  The number of different time orderings is given by
\be
   n_{\rm orderings} = {t \choose s} \quad .         \label{hist_order}
\ee
Combining Eqs.~(\ref{pi}) to (\ref{hist_order}), we obtain
\bea
   p_N^{N_0}(m) &=& {N_0\over N} {t \choose s} {\mathbf{(}N-1\mathbf{)}_{t-s,k} \mathbf{(}1\mathbf{)}_{s,k}\over \mathbf{(}N+k\mathbf{)}_{t,k}} \nonumber \\
                &=& {t \choose s} {\mathbf{(}N-1\mathbf{)}_{t-s,k} \mathbf{(}1\mathbf{)}_{s,k}\over \mathbf{(}N\mathbf{)}_{t,k}}.     \label{ppone}
\eea
This result can be further simplified into beta functions or,  more conveniently,
$k$-beta functions (see e.g.~\cite{ddiaz}),
\be
   B_k(x,y) = {1\over k} B({x\over k},{y\over k})\quad ,
   \nonumber
\ee
giving a remarkably simple final result
\be
   p_N^{N_0}(m) = {t \choose s} {B_k(N_0-m,m)\over B_k(N-1,1)}\quad .       \label{pp1}
\ee

\begin{figure}[t]
\includegraphics[width=\columnwidth]{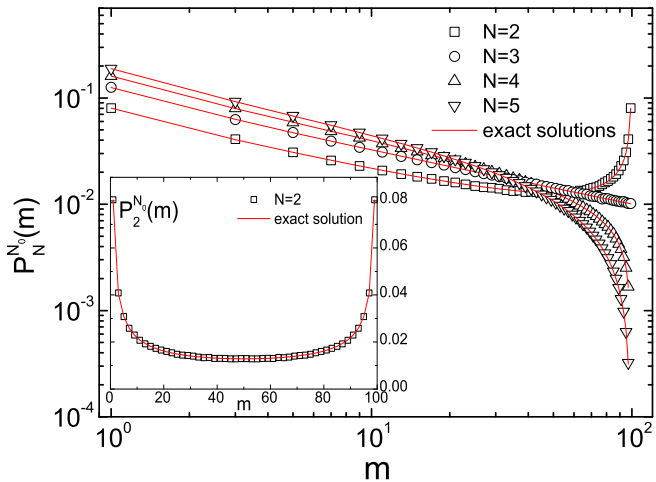}
\vskip -.3cm \caption{(Color online) Cluster size distributions
after $t=50$ events for $k=2$, for different values of $N$
averaged over $10^6$ realizations compared to exact results.  The
large size behavior changes from  increasing to decreasing power law
 at $N=k+1$.  Inset:  The  discrete
arcsine law found for $N=2$. }\label{fig1}
\vskip -0.15in
\end{figure}

We make  a number of observations: (1) For $k=1$ the process maps
to bond percolation on a ring.  For $N=2$, the mass distribution
is uniform over the entire range $m\in [1,N_0-1]$. For $N>2$, the
distribution is proportional to the $(N-2)^{nd}$ factorial power
$((N_0-m-1)(N_0-m-2)\cdots(N_0-m-N+2))$. (2) For $N=2$ and any
$k\geq 1$, $p_N^{N_0}(m)$ is symmetric under the exchange $m
\leftrightarrow N_0-m$. (3) For $N=2$ and $k=2$ we obtain an
equation formally identical to Spitzer's discrete arcsine law for
fluctuations of random walks~\cite{Spitzer}.  (4) Asymptotic power
laws for $N_0\to\infty$ can be determined using Stirling's
formula.  If $N$ is fixed and both $m$ and $(N_0-m) \to\infty$,
\be
    p_N^{N_0}(m) \sim {(t-s)^{\frac{N-1}{k}-1} \over s^{1-\frac{1}{k}}} \quad .
\ee For small masses, this gives a decreasing power law with
exponent $-1+1/k$. For $N = k+1$, the power law $p_N^{N_0}(m) \sim
s^{-1 +1/k}$ holds up to the largest possible value, $m=N_0-N+1$,
and the cutoff is a  step function. For $m/N_0\to 1$ different
power laws appear if $N\neq k+1$, and the sign of the exponent
changes at $N=k+1$. For $N<k+1$, the distribution has a peak at
$m/N_0\to 1$, while it goes to zero for $N > k+1$. These scaling
laws are illustrated for $k=2$ in Fig.~\ref{fig1}. (5)  The
scaling laws found for $m \ll N_0$ are identical to those obtained
by Krapivsky~\cite{Krapivsky} for the well mixed case.
However, the behavior for $m/N_0\to 1$ given in ~\cite{Krapivsky} does not agree with our result.  (6) The probability $p_N^{N_0}(m)$ satisfies a number of recursion relations:
\be
   p_N^{N_0}(m+k) = \frac{m(N_0-m-N+1)}{(m+k-1)(N_0-m-k)}\;  p^{N_0}_N(m)\; ,
   \nonumber
\ee
\be
   p^{N_0}_{N+k}(m) = \frac{N(N_0-m-N+1)}{(N-1)(N_0-N)} \;  p^{N_0}_N (m) \; .\nonumber
\ee
A third {\it nonlinear} recursion relation is given later.

Joint distributions for masses of adjacent clusters can also be found.
We denote by $p_N^{N_0}(m_1,m_2)$ the probability to find
a cluster of mass $m_1$ followed immediately to the right by a cluster
of mass $m_2$. This is non-zero only  if $m_1 = 1+s_1k$ and $m_2=1+s_2k$, where $s_{\alpha}$
is the number of  events needed to form a cluster of mass $m_{\alpha}$. By the
same arguments that led to Eq.~(\ref{ppone}) we get
\be p_N^{N_0}(m_1,m_2) = \binom{t}{s_0, s_1, s_2}
         {\mathbf{(}N-2\mathbf{)}_{s_0,k} \mathbf{(}1 \mathbf{)}_{s_1,k} \mathbf{(}1\mathbf{)}_{s_2,k}\over
          \mathbf{(}N \mathbf{)}_{t,k}}    \label{pp2}
             \; ,
             \nonumber
   \ee
where $s_0 = t-\sum_{\beta=1}^{\alpha} s_{\beta}$ and the first
factor is the multinomial coefficient instead of the binomial
coefficient.
When $\alpha=2$, it is a trinomial coefficient that counts the
number of ways in which the three sequences of events -- for the
two clusters considered, and for all ($N-2$) other clusters -- can
be interleaved in a single history.

For any $1\leq \alpha \leq N-1$ the joint probability distribution
for $\alpha$ consecutive, adjacent clusters is a product of a multinomial coefficient and $\alpha +1$
Pochhammer $k$-symbols, divided by the Pochhammer $k$-symbol related to the total
number of possible histories given $N_0$ initial particles. Defining again $s_0$ as the number of events in all
clusters except the first $\alpha$ ones, we can write the result compactly as
\be
    p_N^{N_0}(m_1,\ldots m_{\alpha}) = \binom{t}{s_0,\ldots
    s_{\alpha}}{(N-\alpha)_{s_0,k}
        \prod_{\beta=1}^{\alpha}(1)_{s_{\beta},k}\over (N)_{t,k}}. \label{ppj}
\ee
In particular, this can also be done for the joint distribution
for all $N$ masses by setting $\alpha=N-1$. The resulting
expression is then manifestly invariant under any permutations of
$N$ numbers $(m_1, \ldots m_N)$. Hence the $N-$ cluster
probability is independent of the spatial ordering of the
clusters. While there are obvious correlations between the mass
values (the sum of all cluster masses must be $N_0$), there are
{\it no spatial correlations}.

We now consider a line of $N_0$ particles with open boundaries.
Again, aggregation events consist of a random choice of a cluster,
followed by its amalgamation with its $k$ nearest neighbors to the
right. The target cluster must be at least $k$ steps away from the
right-most boundary.  Following the same arguments leads
immediately to Eq. (\ref{ppj}) for $\alpha=N-1$, showing that the
two models lead to precisely the same statistics.

The absence of spatial correlations  indicates that the same dynamics might result
for the well-mixed  case.  Now we start with a bucket
containing $N_0$ balls, each of unit mass. An event consists of taking
$k+1$ balls out of the bucket, merging them together, and returning the new ball  to the bucket.
The $k+1$ balls are chosen
completely at random,  independent of  their masses.

The single cluster mass distribution for the well-mixed model can
be obtained using the same strategy as before, but the details are
quite different. Consider  the total number of histories. Since
events now correspond to choosing any $k+1$ balls out of $N_0-kt$
balls, we have, instead of the Pochhammer $k$-symbol, a product of
binomial coefficients, \be
   n_{\rm hist,tot} = {N_0\choose k+1}{N_0-k\choose k+1}\ldots {N+k\choose k+1}.
\ee The expressions for $n_{\rm hist,cluster}$ and $n_{\rm
hist,rest}$ are analogous, with the factors $(m-jk)$ (resp.
$(N_0-m-jk)$) in Eq.~(\ref{hist_clust}) (resp. (\ref{hist_rest}))
replaced by binomial coefficients. The number of time orderings
$n_{\rm orderings}$ is exactly the same as before, but the first
factor $N_0/N$ in Eq.~(\ref{ppone}) has to be replaced by ${1\over
N}{N_0\choose m}$. Putting all these things together, many
cancellations take place, leading exactly to Eq.~(\ref{pp1}). This
argument can be similarly extended to get the full $N$-particle
distribution function, obtaining exactly the same result as
before, for any $k$.

The time-reversed process of aggregation is fragmentation. When
considering the fragmentation process associated with any of these
models,  we have to carefully evaluate fragmentation rates.
Assuming uniform rates would not lead to all time-reversed
histories having the same probability.  Indeed the fraction of all
mergers associated with making a cluster of mass $m'$ is
$(s'/t)=(m'-1)/(N_0 -N)$, which must equal the probability that an
existing cluster of mass $m'$ will fragment at the next step in
the time-reversed process. If it does, then for consistency its
fragmentation products must have a mass distribution given by
$p^{m'}_{k+1}(m)$. A quadratic recursion relation for
$p^{N_0}_{N+k}(m)$ can then be obtained by considering the
likelihood of all fragmentation events in a configuration of $N$
clusters, with $m$ being the mass of one of the resulting $k+1$
fragmentation products.  The relation is \be
  p^{N_0}_{N+k} (m) =
  \sideset{}{'}\sum\limits_{m'=m+k}^{N_0-N+1} {N(m'-1) p^{N_0}_N (m') p^{m'}_{k+1} (m)\over N_0 -N},
              \label{eq8}
              \nonumber
\ee
where the prime on the summation symbol indicates that $m'$ must
increase in steps of $k$.


In summary, we derived complete  solutions for the probability to
find any given state in three models -- well-mixed solutions,
particles on a ring reacting with their $k$ nearest neighbors, and
the same reaction for particles on a line with open boundaries --
and show that these solutions are precisely the same. The fact
that we could solve exactly a one dimensional model without
detailed balance might seem surprising  since such models are in
general
not solvable.   
It stems from the fact that spatial correlations, although
 {\it a priori} not  excluded, are in fact absent.
Related to this is our finding that the well-mixed models have
exactly the same solutions.  Our method can be used to
solve the model where the target cluster is picked with a
probability proportional to its mass~\cite{SWSon2011}.  Perhaps generalizations of these observations
hold true for more complicated models, in which case weighted path integrals would replace sums over histories.

We have pointed out a direct mapping between irreversible aggregation and RSR.
The latter was motivated by claims that one can
define finite fractal dimensions for real
networks~\cite{CSong2005}, using similar but more complicated and
ambiguous schemes. Results for RSR with $\ell=1$ on various graphs
(critical trees~\cite{bizhani2010}, Erd\"os-Renyi and
Barabasi-Albert networks~\cite{bizhani2011}, and regular
lattices~\cite{christensen}) concur with our present conclusions
for $k=2$.   Apart from studying a system that is sufficiently
simple to be exactly solvable and that is obviously not fractal,
here we presented results for
$\ell>1$, showing that scaling laws depend in a
non-trivial way on $\ell$.  Results for the elementary network (a one dimensional line)  examined analytically here proves that scaling under stochastic network renormalization arises from an underlying percolation transition in aggregation and does not prove fractality or self-similarity of the underlying graph.

Our mapping suggests that the critical behavior of aggregation may
also turn up in ``Graphity"~\cite{Konopka2006,Hamma2010}  or
related models, where geometry, gravity, and matter emerge through
an aggregation process of an underlying graph.  ``Geometrogenesis"
is the complementary process of infinite cluster formation in
irreversible aggregation.  In that case, scaling that depends on
the microscopic coarse-graining scale $\ell$ seems to add further
obstacles to the persistent and challenging problem to derive a
large scale theory of gravity from microscopic graph models.

\bibliography{coagulation_references}
\end{document}